\documentclass[10pt]{article}
\usepackage[OE]{express}
\usepackage{bbold}
\usepackage{amsmath}

\begin{document}
\title{Analytical mode normalization and resonant state expansion for optical fibers - an efficient tool to model transverse disorder}

\author{S.~Upendar,\authormark{1} I.~Allayarov,\authormark{1} M.~A.~Schmidt,\authormark{2,3} and T.~Weiss\authormark{1}}

\address{\authormark{1} $\mathrm{4^{th}}$ Physics Institute and Research Center SCoPE, University of Stuttgart, Pfaffenwaldring 57, 70569 Stuttgart, Germany \newline
\authormark{2} Leibniz Institute of Photonic Technology e.V. Albert-Einstein-Str. 9, 07745, Jena, Germany \newline
\authormark{3} Otto Schott
Institute of Material Research, Friedrich Schiller University, Faunhoferstr. 6, 07743, Jena, Germany}

\email{\authormark{*}s.upendar@pi4.uni-stuttgart.de} 



\begin{abstract}
We adapt the resonant state expansion to optical fibers such as capillary and photonic crystal fibers. As a key requirement of the resonant state expansion and any related perturbative approach, we derive the correct analytical normalization for all modes of these fiber structures, including leaky modes that radiate energy perpendicular to the direction of propagation and have fields that grow with distance from the fiber core. Based on the normalized fiber modes, an eigenvalue equation is derived that allows for calculating the influence of small and large perturbations such as structural disorder on the guiding properties. This is demonstrated for two test systems: a capillary fiber and an endlessly single mode fiber.
\end{abstract}

\ocis{(060.5295) Photonic crystal fibers; (060.4005) Micro structured fibers; (080.1753) Computation Methods;}


\section{Introduction}

Photonic crystal fibers guide light in a central defect core surrounded by a periodic cladding\cite{russell2003photonic}. The guiding mechanism of the photonic crystal fiber can be a bandgap effect or modified total internal reflection in cases where the index of the core is larger than the effective cladding index. These fibers feature a high degree of light confinement, highly tunable dispersion properties\cite{knight2000anomalous}, and single mode operation\cite{knight1996all}. Photonic crystal fibers are extensively used in gas sensing\cite{ritari2004gas}, nonlinear optics such as supercontinuum generation\cite{dudley2006supercontinuum}, and many more applications\cite{poli2007photonic,humbert2004hollow,benabid2002stimulated,li2017guiding,granzow2011bandgap,schmidt2009all}.  

In theoretical investigations, an ideal cladding is usually used to analyze such structures, while a fabricated photonic crystal fiber cladding is never truly perfect\cite{frosz2013five,roberts2005ultimate}. The fabrication process itself gives rise to shape and position disorders that influence the guiding properties. Studying that influence requires investigating many realizations\cite{nau2006disorder}, which is rather tedious in conventional numerical approaches. In contrast, the resonant state expansion has proven rather efficient for investigating a large set of similar three-dimensional resonator systems \cite{doost2014resonant,muljarov2011brillouin,doost2013resonant,muljarov2016exact} and slab waveguides \cite{armitage2014resonant,lobanov2017resonant}. The resonant state expansion is a rigorous perturbative approach, in which the resonant states (also known as quasi-normal modes \cite{sauvan2013theory,kristensen2013modes}) of a reference system are used to setup an eigenvalue equation that provides the resonant states of a perturbed system. Here, we adapt the resonant state expansion to fiber geometries, in which the core and cladding modes constitute the resonant states, and treat disorder as a perturbation of the perfect cladding system. 

\begin{figure}[htbp]
 \centering
  \includegraphics[width=10cm]{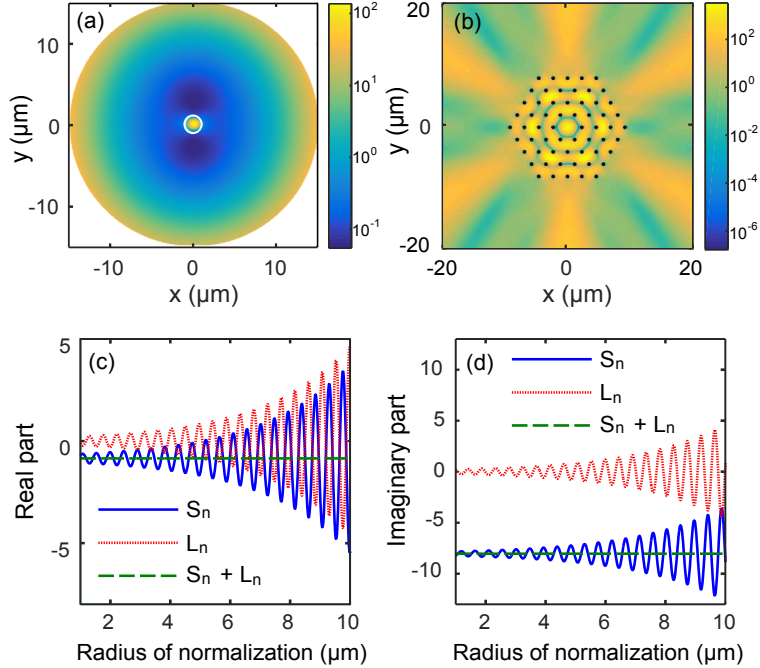}
 \caption{(a) Axial component of the time-averaged Poynting vector of the fundamental core mode of a step index fiber with refractive indices of $1$ and $1.44$ in the core and cladding region, respectively, and a core radius of \mbox{$1$ \textmu m} at a wavelength of \mbox{$1$ \textmu m}. (b) Axial component of the time-averaged Poynting vector for a higher-order core mode of a silica-air photonic crystal fiber with four rings of air holes of radius \mbox{$0.25$ \textmu m} and pitch \mbox{$2.3$ \textmu m} around a single-defect core. The refractive index of silica is taken as $1.44$. The considered wavelength is \mbox{$1$ \textmu m}. Both modes in (a) and (b) exhibit fields that grow in the exterior with distance from the core. Panels (c) and (d) depict the real and imaginary parts of the surface term (blue solid line) and line term (red dotted line) of the normalization Eq.~(\ref{Nm}) as a function of the radius of normalization. Evidently, the divergence of the fields is manifested in the surface and line terms, while it is countervailed in their sum as the normalization constant.}
\end{figure}

As in any perturbation theory, the key factor in the resonant state expansion is the normalization of the resonant states. The normalization is not trivial, since the solutions of Maxwell's equations include leaky modes\cite{sammut1976leaky1}. These modes radiate energy perpendicular to the fiber axis and have fields that grow with distance from the fiber core. This is displayed in Fig.~1 for a capillary fiber with air core and silica cladding (a) and a photonic crystal fiber with air inclusions and silica background (b). A lot of work has been devoted to the normalization of leaky modes\cite{snyder2012optical,marcuse1974theory,lee1995leaky,lai1990time}. The most sophisticated approach is introducing a complex coordinate transformation in the exterior that suppresses the growth\cite{sammut1976leaky}, which is equivalent to using perfectly matched layers and extending the area of normalization to the perfectly matched layers\cite{sauvan2013theory}. In contrast, we derive here an analytical normalization that can be calculated without any perfectly matched layers and is valid for both guided as well as leaky modes. Our new normalization can be easily applied when using standard numerical methods for the calculation of modes.

Here, the properties of the resonant state expansion with our analytical mode normalization is demonstrated for two fiber geometries. In the first example, we use the analytical solutions for a capillary fiber as basis to model the influence of a homogeneous change of the refractive index of the fiber core on the propagation constants of the fiber modes. In the second example, we investigate the influence of diameter disorder on the modal properties of an endlessly single mode fiber\cite{birks1997endlessly}.

\section{Theory}

Maxwell's Equations can be summarized in real space and frequency domain with time dependence exp$(-i\omega t)$ by the compact operator form\cite{muljarov2018resonant}

\begin{equation}
\underbrace{
\begin{pmatrix}
k_0\varepsilon & -\nabla\times \\ -\nabla\times & k_0\mu
\end{pmatrix}}_{\equiv \varmathbb{M}_0}
\underbrace{
\begin{pmatrix}
\mathbf{E} \\ i\mathbf{H}
\end{pmatrix}}_{\equiv \mathbf{\varmathbb{F}}}  
=
\underbrace{
\begin{pmatrix}
\mathbf{J}_E \\ i\mathbf{J}_H
\end{pmatrix}}_{\equiv \mathbf{\varmathbb{J}}},
\label{MxEsource}
\end{equation}
with electric and magnetic fields \textbf{E} and \textbf{H}, respectively, permittivity and permeability tensors $\varepsilon$ and $\mu$, respectively, and $k_0 = \omega/c$. The right-hand side contains the electric source term $\mathbf{J}_E=-4\pi i \mathbf{j}/c$ with current density $\mathbf{j}$, and  the magnetic source term $\mathbf{J}_H$ that has been introduced for the sake of symmetry.

For optical fibers, the permittivity and permeability tensors are translationally symmetric along the direction of propagation, which we choose as the $z$ direction of our coordinate system. Defining the Fourier transform in this direction as
\begin{equation}
\hat{f}(\textbf{r}_\|;\beta) = \frac{1}{2\pi}\int\limits_{-\infty}^\infty \mathrm{d}z f(\textbf{r}_\|;z)e^{-i\beta z},
\end{equation}
with $\mathbf{r}_{||}$ being the projection of $\mathbf{r}$ to the $xy$ plane and the hat denoting Fourier transformed quantities, the Fourier transform of Eq.~(\ref{MxEsource}) yields

\begin{equation}
\begin{pmatrix}
k_0\varepsilon & -\hat{\nabla}_\beta\times \\ -\hat{\nabla}_\beta\times & k_0\mu
\end{pmatrix}
\begin{pmatrix}
\hat{\textbf{E}} \\ i\hat{\textbf{H}}
\end{pmatrix}
= 
\begin{pmatrix}
\hat{\textbf{J}}_E \\ i\hat{\textbf{J}}_H
\end{pmatrix}, \hspace{2mm} \mathrm{with} \hspace{2mm} \hat{\nabla}_\beta \equiv 
\begin{pmatrix}
\partial_x \\ \partial_y \\ i\beta 
\label{MxEsourceJ}
\end{pmatrix}.
\end{equation}

The Green's dyadic\cite{tai1994dyadic} of Eq.~(\ref{MxEsourceJ}) satisfies the relation
\begin{equation}
\hat{\varmathbb{M}}_0(\textbf{r}_\|;\beta)\hat{\varmathbb{G}}(\textbf{r}_\|,\textbf{r}'_\|;\beta) = \mathbb{1}\delta(\textbf{r}_\| - \textbf{r}'_\|),
\end{equation}
and provides the solutions $\hat{\varmathbb{F}}$ of Eq.~(\ref{MxEsourceJ}) for a given source $\hat{\varmathbb{J}}$ as

\begin{equation}
\hat{\varmathbb{F}}(\textbf{r}_\|) = \int \mathrm{d}\textbf{r}'_\| \hspace{1mm}\hat{\varmathbb{G}}(\textbf{r}_\|,\textbf{r}'_\|; \beta)\hat{\varmathbb{J}}(\textbf{r}'_\|).
\label{gfsource}
\end{equation}
The Green's dyadic can be expanded in terms of the resonant states \cite{muljarov2011brillouin,doost2013resonant,weiss2016dark,muljarov2018resonant,muljarov2016resonant,doost2012resonant,doost2014resonant}, which are solutions of Eq.~(\ref{MxEsourceJ}) in the absence of sources for outgoing boundary conditions with eigenvectors $\hat{\varmathbb{F}}_n$ and eigenvalues $\beta_n$:
\begin{equation}
\hat{\varmathbb{M}}_0(\mathbf{r}_{||};\beta_n) \hat{\varmathbb{F}}_n = 0.
\label{MxE}
\end{equation}
Using the Mittag-Leffler theorem\cite{arfken1999mathematical} and the reciprocity principle\cite{weiss2017analytical}, it follows that
\begin{equation}
\hat{\varmathbb{G}}(\textbf{r}_\|,\textbf{r}'_\|,\beta) =-\sum_n \frac{\hat{\varmathbb{F}}_n(\textbf{r}_\|)\otimes\hat{\varmathbb{F}}^{\mathrm{R}}_n(\textbf{r}'_\|)}{2N_n(\beta-\beta_n)} + \Delta\hat{\varmathbb{G}}_\mathrm{cuts},
\label{grf}
\end{equation}
with $\otimes$ denoting the outer vector product, and $N_n$ being the normalization constant in order to assign the appropriate weight to the resonant states, since Eq.~(\ref{MxE}) provides the resonant field distribtuions only up to a constant factor. The factor $-1/2$ has been introduced for later convenience. The superscript $\mathrm{R}$ denotes the reciprocal conjugate resonant state, which is a solution of Eq.~(\ref{MxE}) at $-\beta_n$. Note that Eq.~(\ref{grf}) is only valid within the regions of spatial inhomogeneities of the fiber \cite{weiss2017analytical}, where the leaky modes do not exhibit any growth. Furthermore, $\Delta\hat{\varmathbb{G}}_\mathrm{cuts}$ denotes cut contributions due to branch cuts in the involved analytical functions. In the following, we will focus on the contribution of the resonant states, keeping in mind that we can treat the cut contributions in a similar manner in numerical calculations\cite{doost2013resonant,weiss2016dark}. 

The derivation of the normalization constant is described in detail in Appendix A. The resulting normalization can be split into two terms comprising of a surface and a line integral that are evaluated on a circle with radius $R$ outside the region of inhomogeneities, which yields

\begin{equation}
N_n = S_n + L_n,
\label{Nm}
\end{equation}
with the surface term
\begin{equation}
S_n =  \int\limits_0^R \rho \mathrm{d}\rho \int\limits_0^{2\pi} \mathrm{d}\phi\hspace{1mm} (\hat{E}_{n,\rho}\hat{H}_{n,\phi} - \hat{E}_{n,\phi}\hat{H}_{n,\rho}),
\label{Sn}
\end{equation}
which is proportional to the integral over the $z$ component of the real-valued Poynting vector, and the line term

\begin{equation}
L_n = \frac{\varepsilon\mu k_0^2 + \beta_n^2}{2\varkappa_n^4}\int\limits_0^{2\pi} \mathrm{d}\phi\hspace{1mm}\bigg(\hat{E}_{n,z}\frac{\partial \hat{H}_{n,z}}{\partial \phi} - \hat{H}_{n,z}\frac{\partial \hat{E}_{n,z}}{\partial \phi}\bigg)_R \nonumber
\end{equation}
\begin{equation}
+ \frac{k_0\beta_n\rho^2}{2\varkappa_n^4}\int\limits_0^{2\pi} \mathrm{d}\phi\hspace{1mm} \bigg\{\hspace{0.5mm}\mu\bigg[\bigg(\frac{\partial \hat{H}_{n,z}}{\partial \rho}\bigg)^2 - \rho\hat{H}_{n,z}\frac{\partial}{\partial \rho}\bigg(\frac{1}{\rho}\frac{\partial \hat{H}_{n,z} }{\partial \rho}\bigg)\bigg] 
+ \varepsilon\bigg[\bigg(\frac{\partial \hat{E}_{n,z}}{\partial \rho}\bigg)^2 - \rho\hat{E}_{n,z}\frac{\partial}{\partial \rho}\bigg(\frac{1}{\rho}\frac{\partial \hat{E}_{n,z} }{\partial \rho}\bigg)\bigg]\hspace{1mm}\bigg\}_R, 
\label{Ln}
\end{equation}
where the subscript $R$ indicates that the integrand is evaluated at radius $R$, and
\begin{equation}
\varkappa_n^2 = \varepsilon\mu k_0^2 - \beta_n^2.
\label{kappa_rel}
\end{equation} 

For truly guided modes, the resonant states decay outside the regions of spatial inhomogeneities, so that the line term vanishes in the limit of $R\rightarrow\infty$. This results in the rather well-known normalization of resonant states by the integral over the $z$ component of the Poynting vector \cite{sammut1976leaky}. For leaky modes, both the line and the surface term diverge. However, their sum countervails this divergence, resulting in a normalization constant independent of the radius of normalization, see Fig. 1. Hence, it is possible to calculate the normalization constant for a small area surrounding the regions of spatial inhomogeneities, without the need of including perfectly matched layers \cite{sauvan2013theory} or, equivalently, complex coordinates \cite{sammut1976leaky}. Furthermore, it should be noted that this approach also simplifies the normalization of truly guided modes in numerical calculations, since it allows to restrict the normalization integrals, and, thus, the computational domain, to a small area. 

Using the normalization to gauge the correct weight of the resonances, it is possible to determine the resonant states of a perturbed system (denoted by subscript $\nu$) with perturbation $\Delta\varepsilon$ and $\Delta\mu$ that exhibits the same translational symmetry as $\varepsilon$ and $\mu$ and vanish outside the regions of spatial ingomogeneities. The Maxwell operator $\hat{\varmathbb{M}}$ of the perturbed system can be separated into the operator $\hat{\varmathbb{M}}_0$ of the unperturbed system and the deviation $\Delta \hat{\varmathbb{M}}$ as $\hat{\varmathbb{M}}=\hat{\varmathbb{M}}_0+\Delta\hat{\varmathbb{M}}$, with

\begin{equation}
\Delta\hat{\varmathbb{M}} = 
\begin{pmatrix}
k_0\Delta\varepsilon & 0 \\0 & k_0\Delta\mu
\end{pmatrix}.
\end{equation}
Thus, we can recast Eq.~(\ref{MxE}) in the form 
\begin{equation}
\hat{\varmathbb{M}}_0 \hat{\varmathbb{F}}_\nu = -\Delta \hat{\varmathbb{M}} \hat{\varmathbb{F}}_\nu.
\end{equation}
Using Eq.~(\ref{gfsource}), we therefore obtain

\begin{equation}
\hat{\mathbf{\varmathbb{F}}}_\nu(\textbf{r}_\|) = -\int \mathrm{d}\textbf{r}'_\|\hspace{1mm} \hat{\varmathbb{G}}(\textbf{r}_\|,\textbf{r}'_\|,\beta_\nu) \Delta\hat{\varmathbb{M}}\hat{\mathbf{\varmathbb{F}}}_\nu(\textbf{r}'_\|) .
\label{greensdydicdeltaM}
\end{equation}
Next, we construct the resonant states of the perturbed system as a linear combination of the normalized resonant states of the unperturbed system:
\begin{equation}
\hat{\mathbf{\varmathbb{F}}}_\nu = \sum_n b_n\hat{\mathbf{\varmathbb{F}}}_n,
\label{expansion}
\end{equation}
Using this ansatz in Eq.~(\ref{greensdydicdeltaM}) and equating it for each $\hat{\mathbf{\varmathbb{F}}}_n$ independently, we obtain 
\begin{equation}
\beta_\nu b_n = \beta_n b_n + \frac{1}{2}\sum_{n'} V_{nn'}b_{n'},
\label{eig}
\end{equation}
where
\begin{equation}
V_{nn'} = \int \mathrm{d}\textbf{r}_\|\hspace{1mm}\hat{\mathbf{\varmathbb{F}}}^\mathrm{R}_n(\textbf{r}_\|)\cdot\Delta\varmathbb{M}\hat{\mathbf{\varmathbb{F}}}_{n'}(\textbf{r}_\|).
\end{equation}
The above equations describe a linear eigenvalue problem with $\beta_\nu$ as the eigenvalue. Note that the sum in Eq.~(\ref{expansion}) is carried out over all resonant states of the unperturbed system, but in real calculations, a truncated basis is used to expand $\hat{\mathbf{\varmathbb{F}}}_\nu$. The choice of the basis size has to be taken large enough to accurately account for the perturbations in the system. 

\section{Results and discussion}

We first consider as our unperturbed system a capillary fiber with core index $1$ and cladding index $1.44$ having a core radius of \mbox{$8$ \textmu m}. The values of the propagation constant, and hence, the effective index of the fundamental HE$_{11}$ mode as well as those of higher-order modes have been determined analytically by solving their characteristic equation \cite{snyder2012optical,jackson1999classical,marcatili1964hollow} at a wavelength of $1$ \textmu m. The fields of the fiber are proportional to Bessel functions inside the core and outgoing Hankel functions in the cladding region. 

\begin{figure}[htbp]
 \centering
  \includegraphics[width=10cm]{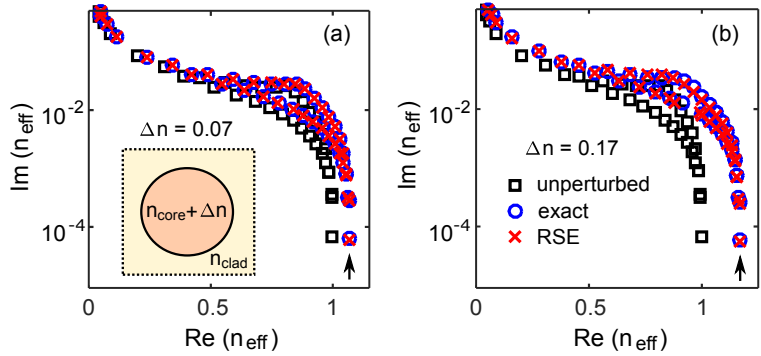}
 \caption{Effective refractive indices of modes in a capillary fiber with a homogeneous perturbation in the core region of (a) $\Delta n = 0.07 $ and (b) $\Delta n = 0.17 $. The results from the resonant state expansion (red crosses) are compared with the exact analytical solution (blue circles) for the perturbed system at a wavelength of $1$ \textmu m. The unperturbed system has a core index of 1, cladding index of 1.44, and a radius of \mbox{$8$ \textmu m}, with its effective refractive indices denoted by black squares. The number of modes used is 154. The black arrow indicates the fundamental core mode. }
\end{figure}

A homogeneous perturbation of $ \Delta n $ is introduced inside the core of the fiber changing the core index to \mbox{n\textsubscript{core}$ + \Delta n$}. As our perturbation is azimuthally symmetric, we only require modes of the same symmetry as the fundamental core mode to set up our eigenvalue problem of Eq.~(\ref{eig}). The comparison of the resonant state expansion with the exact analytical solution for the fundamental and higher order modes of azimuthal order $m=1$ is shown in Fig. 2 for (a) $\Delta n = 0.07$ and (b) $\Delta n = 0.17$. We can see that there is a good agreement not only for the fundamental mode (indicated by the arrow) but also for the higher order modes of the system. The number of modes used is 154, with pairs of modes with propagation constants $\beta_n$ and $-\beta_n$. The relative error given by $|1- n_{\mathrm{eff}}^{\mathrm{RSE}}/n_{\mathrm{eff}}^{\mathrm{exact}}|$ is on the order of $10^{-6}$ for the fundamental mode for a core perturbation of  $\Delta n = 0.17$. The average relative error of the higher-order modes is on the order of $10^{-3}$ for the same perturbation.

\begin{figure}[htbp]
 \centering
  \includegraphics[width=10cm]{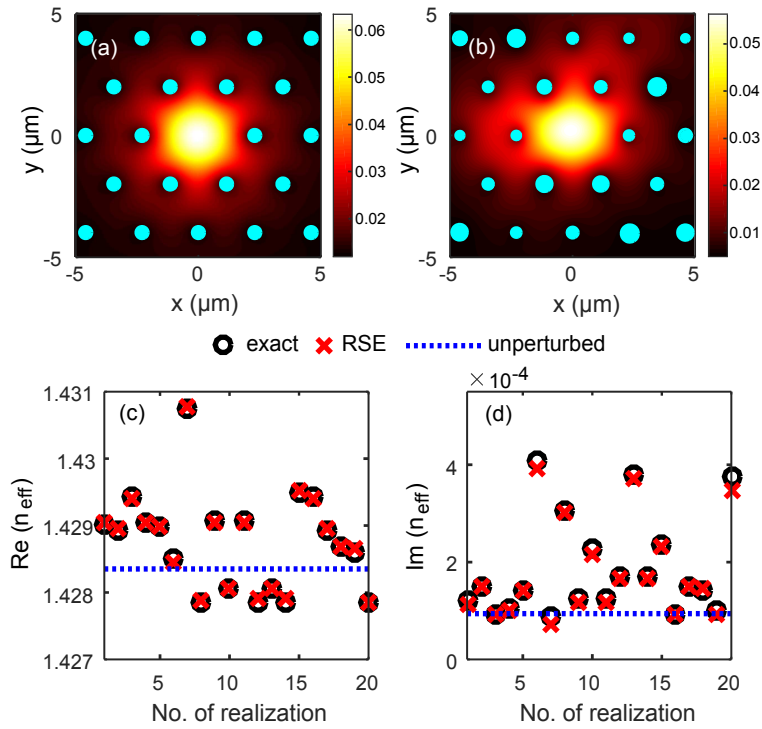}
 \caption{Axial component of the Poynting vector of the fundamental core mode of a silica-air photonic crystal fiber with diameter disorder for disorder parameter (a) $\Delta = 0$ \textmu m and (b) \mbox{$\Delta= 0.1$ \textmu m}. The disorder parameter provides the range of radii in the disordered fiber as $r_0 \pm \Delta$, with $r_0$ being the radius of the air holes in the ordered fiber. The geometrical parameters of the fiber are the same as in Fig. 1(b). Panels (c) and (d) show the comparison of the real and imaginary parts of the effective indices from the resonant state expansion (red crosses) with the exact numerical solution of the perturbed system (black circles) for 20 realizations of disorder at a wavelength of \mbox{$1.55$ \textmu m}. The number of modes used for the resonant state expansion is 190. The blue dotted line indicates the effective index for an unperturbed cladding. }
\end{figure}

As a second example, we consider a silica-air photonic crystal fiber of air holes with radius \mbox{$r_0 = 0.25$ \textmu m} in four cladding rings with pitch \mbox{$2.3$ \textmu m} around a single defect core. We numerically solve for modes of the fiber \cite{kuhlmeycomputer,kuhlmey2002multipole,white2002multipole1,fini2004improved} with a perfect cladding structure and use them as basis for a perturbed system, in which we introduce diameter disorder in each and every inclusion in the cladding region. The range of the diameter disorder is determined by the disorder parameter $\Delta$ as $r_0 \pm \Delta$. Within that radius range of width $2\Delta$, a uniform distribution of disorder is used. The probability density for a uniform distribution is given as,
\begin{equation}
f(r) = \begin{cases}
       \frac{1}{2\Delta}\hspace{2mm}\mathrm{for}\hspace{2mm} r_0 - \Delta \leq r \leq r_0 + \Delta \\
       0 \hspace{2mm}\mathrm{for}\hspace{2mm} r < r_0 - \Delta \hspace{1mm}\mathrm{or}\hspace{1mm} r > r_0 + \Delta
       \end{cases}
\end{equation}
 
We set up our eigenvalue problem with $190$ modes with effective indices relatively close to the fundamental mode. The comparison of the real and imaginary part of the effective index obtained from the resonant state expansion (red crosses) and full numerical calculations (black circles) can be seen in Fig. 3 (c) and (d), respectively, for 20 realizations and $\Delta = 0.1$ \textmu m at a wavelength of $1.55$ \textmu m. Evidently, there is a good agreement between the two methods for the shown realizations.  

In Fig. 4 (a) and (b), we display the real and imaginary parts of the effective index averaged over 200 realizations for disorder parameters ranging from $\Delta = 0$ to $0.11$ \textmu m. More specifically, we generate 200 sets of random numbers between 0 and 1 for each air hole and multiply them with different values of $\Delta$ in order to generate the disordered fibers. The standard deviation of the effective index is plotted as error bars that grow with increasing $\Delta$. Interestingly, the average \mbox{Re(n\textsubscript{eff})} has a linear dependence with $\Delta$ while the Im(n\textsubscript{eff}) exhibits a more quadratic behavior.

\begin{figure}[htbp]
 \centering
  \includegraphics[width=11cm]{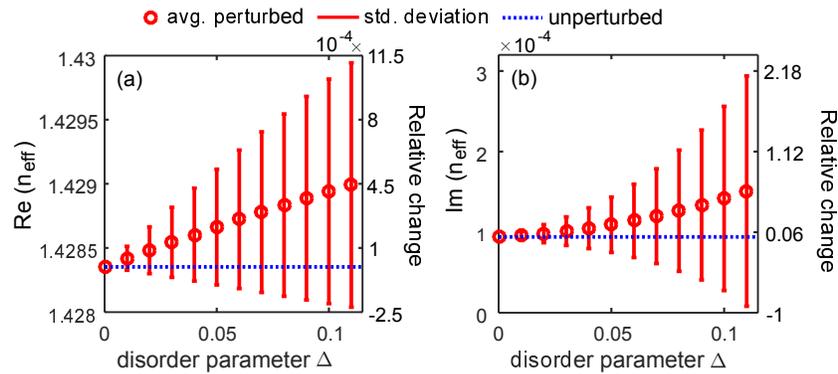}
 \caption{Real (a) and imaginary (b) part of the effective index of the fundamental core mode as a function of the disorder parameter $\Delta $ averaged over 200 realizations of diameter disorder at a wavelength of $1.55$ \textmu m. The averaged real part grows almost linearly with increasing $\Delta$, while the imaginary part is growing quadratically. The standard deviation is indicated by the errorbars. The blue dotted line indicates the effective index of the unperturbed cladding.}
\end{figure}

\section{Conclusion}

We have derived an analytical normalization for modes in fiber geometries that is valid not only for guided but also for leaky modes. We have shown that the normalization constant is independent of the radius of integration even for leaky modes with fields that grow with distance from the fiber core. Thus, it is possible to set up an eigenvalue equation that allows us to calculate the effective refractive indices of modes in a perturbed system. The accuracy of this so-called resonant state expansion has been demonstrated for capillary-type and photonic crystal fibers. For the latter, we have studied diameter disorder in the cladding of a silica-air photonic crystal fiber for different disorder parameters averaged over many realizations. Here, the resonant state expansion is clearly superior compared to full numerical simulations, since it does not require to repeatedly solve Maxwell's equations, while the numerical effort for solving the eigenvalue equation is rather low. Thus, it is possible to derive the influence of disorder on the guiding properties such as propagation constant and loss efficiently.  

\appendix
\section{Normalization}

Let us consider the Maxwell's equation with a source term that vanishes at resonance:

\begin{equation}
\hat{\varmathbb{M}}_0(\textbf{r}_\|;\beta)\hat{\mathbf{\varmathbb{F}}} = (\beta - \beta_n)\sigma_n(\textbf{r}_\|).
\label{sourceatres}
\end{equation}
Here, $\sigma_n(\textbf{r}_\|)$ is chosen to vanish outside the region of spatial inhomogeneities. Taking the source term and convoluting with the Green's dyadic in the limit $\beta \rightarrow \beta_n$, we get

\begin{equation}
\hat{\varmathbb{F}}_n(\textbf{r}_\|) = \lim_{\beta \to \beta_n}\sum_{n'}\frac{-1}{2N_{n'}}\frac{\beta - \beta_n}{\beta - \beta_{n'}}\hspace{0.5mm}\hat{\varmathbb{F}}_{n'}(\textbf{r}_\|)\int \mathrm{d}\textbf{r}_\|'\hspace{1mm}\hat{\varmathbb{F}}^{\mathrm{R}}_{n'}(\textbf{r}'_\|)\sigma_n(\textbf{r}'_\|).
\end{equation}
This can be only fulfilled for 

\begin{equation}
\int \mathrm{d}\textbf{r}_\|'\hspace{1mm}\hat{\mathbf{\varmathbb{F}}}^\mathrm{R}_n(\mathbf{r}'_\|)\sigma_n(\textbf{r}'_\|) = -2N_n.
\end{equation}
To derive the normalization equation, we multiply Eq. (\ref{sourceatres}) with $\hat{\mathbf{\varmathbb{F}}}^\mathrm{R}_n(\textbf{r}_\|)$ and subtract a zero in the form of

\begin{equation}
0 = \hat{\mathbf{\varmathbb{F}}}(\textbf{r}_\|;\beta)\cdot
\hat{\varmathbb{M}}_0(\textbf{r}_\|;-\beta_n)\hat{\mathbf{\varmathbb{F}}}^\mathrm{R}_n(\textbf{r}_\|),
\end{equation}
to obtain, 

\begin{equation}
\hat{\mathbf{\varmathbb{F}}}^\mathrm{R}_n(\textbf{r}_\|)\cdot\hat{\varmathbb{M}}_0(\textbf{r}_\|;\beta)\hat{\mathbf{\varmathbb{F}}}(\textbf{r}_\|;\beta) - \hat{\mathbf{\varmathbb{F}}}(\textbf{r}_\|;\beta)\cdot\hat{\varmathbb{M}}_0(\textbf{r}_\|;-\beta_n)\hat{\mathbf{\varmathbb{F}}}^\mathrm{R}_n(\textbf{r}_\|) = (\beta - \beta_n)\hat{\mathbf{\varmathbb{F}}}^\mathrm{R}_n(\textbf{r}_\|)\cdot\sigma_n(\textbf{r}_\|).
\end{equation}
Dividing by $\beta - \beta_n$ and integrating over the spatial inhomogeneities in the limit $\beta \to \beta_n$, we get

\begin{equation}
-2N_n = \lim_{\beta \to \beta_n}\int{\mathrm{d}\textbf{r}_\|\hspace{1mm}\frac{-i}{\beta - \beta_n}\nabla_\|\cdot[\hat{\textbf{E}}(\textbf{r}_\|;\beta)\times\hat{\textbf{H}}_n^\mathrm{R}(\textbf{r}_\|) 
-\hat{\textbf{E}}^\mathrm{R}_n(\textbf{r}_\|)\times\hat{\textbf{H}}(\textbf{r}_\|;\beta)]} \nonumber
\end{equation}

\begin{equation}
+ \int \mathrm{d}\textbf{r}_\|\hspace{1mm}[\hat{\textbf{E}}_n(\textbf{r}_\|)\times\hat{\textbf{H}}_n^\mathrm{R}(\textbf{r}_\|)-\hat{\textbf{E}}_n^\mathrm{R}(\textbf{r}_\|)\times\hat{\textbf{H}}_n(\textbf{r}_\|)]_z.
\label{Nm_term}
\end{equation}
The subscript $z$ indicates the integration of the $z$ component in the second term which results in the surface integral of Eq.~(\ref{Sn}) when using that, due to symmetry, the in-plane components of the electric field and the $z$ component of the magnetic field of resonant states with eigenvalues $\beta_n$ and $-\beta_n$ are identical, while we have to multiply all other components with $-1$ in order to convert $\hat{\varmathbb{F}}_n^\mathrm{R}$ into $\hat{\varmathbb{F}}_n$. The first term can be converted to a line integral by using the divergence theorem. The curve of integration is taken as a circle of radius $R$ outside the region of inhomogeneities. For evaluating the limit $\beta \to \beta_n$, we carry out a Taylor expansion around $\beta_n$ as
\begin{equation}
\hat{\mathbf{\varmathbb{F}}}(\textbf{r}_\|;\beta) = \hat{\mathbf{\varmathbb{F}}}_n(\textbf{r}_\|) + \frac{(\beta - \beta_n)}{1!}\frac{\partial\hat{\mathbf{\varmathbb{F}}}}{\partial \beta}\bigg\vert_{\hspace{0.5mm}\beta_n} + \frac{(\beta - \beta_n)^2}{2!}\frac{\partial^2\hat{\mathbf{\varmathbb{F}}}}{\partial \beta^2}\bigg\vert_{\hspace{0.5mm}\beta_n}+...\hspace{1mm},
\label{taylor}
\end{equation}
which results in a line integral that contains $\hat{\mathbf{\varmathbb{F}}}^\mathrm{R}_n$ as well as first-order derivatives of $\hat{\mathbf{\varmathbb{F}}}$ with respect to $\beta$ at $\beta_n$. Moreover, due to the aforementioned relations between $\hat{\mathbf{\varmathbb{F}}}^\mathrm{R}_n$ and $\hat{\mathbf{\varmathbb{F}}}_n$, we can rewrite Eq.~(\ref{Nm_term}) as
\begin{equation}
N_n = \frac{\beta_n\rho}{2i\varkappa_n}\int\limits_0^{2\pi}\mathrm{d}\phi\hspace{1mm}(\frac{\partial\hat{E}_{n,\phi}}{\partial \varkappa}\hat{H}_{n,z} 
                                                     +\frac{\partial\hat{E}_{n,z}}{\partial \varkappa}\hat{H}_{n,\phi} 
																										- \frac{\partial\hat{H}_{n,\phi}}{\partial \varkappa}\hat{E}_{n,z}
																										- \frac{\partial\hat{H}_{n,z}}{\partial \varkappa}\hat{E}_{n,\phi}) \nonumber
\end{equation}
\begin{equation}
+ \int \mathrm{d}\textbf{r}_\|\hspace{1mm}(\hat{E}_{n,\rho}\hat{H}_{n,\phi} - \hat{E}_{n,\phi}\hat{H}_{n,\rho}).
\end{equation}

Note that we have converted the derivative with respect to $\beta$ to a derivative with respect to $\varkappa$ by using the relation in Eq.~(\ref{kappa_rel}). The derivative with respect to $\varkappa$ can then be converted to spatial derivatives by using the following relations:

\begin{equation}
\frac{\partial \hat{E}_z}{\partial \varkappa} = \frac{\rho}{\varkappa}\frac{\partial  \hat{E}_z}{\partial\rho},
\hspace{5mm}\frac{\partial  \hat{H}_z}{\partial \varkappa} = \frac{\rho}{\varkappa}\frac{\partial \hat{H}_z}{\partial\rho}.
\label{ezhztransformation}
\end{equation}
The $\hat{E}_\phi$ and $\hat{H}_\phi$ field components can be derived from the $\hat{E}_z$ and $\hat{H}_z$ field components as
\begin{equation}
\hat{E}_\phi = \frac{i\beta}{\varkappa^2\rho}\frac{\partial \hat{E}_z}{\partial \phi} - \frac{ik_0\mu}{\varkappa^2}\frac{\partial \hat{H}_z}{\partial \rho},
\hspace{5mm}\hat{H}_\phi = \frac{i\beta}{\varkappa^2\rho}\frac{\partial \hat{H}_z}{\partial \phi} + \frac{ik_0\varepsilon}{\varkappa^2}\frac{\partial \hat{E}_z}{\partial \rho},
\label{ephihphi}
\end{equation}
and they can be differentiated with respect to $\varkappa$ by using the relations for $\hat{E}_z$ and $\hat{H}_z$ given in Eq.~(\ref{ezhztransformation}). Substituting $\hat{E}_\phi$ and $\hat{H}_\phi$ by Eq.~(\ref{ephihphi}) and using that 
\begin{equation}
\int\limits_0^{2\pi}\mathrm{d}\phi\hspace{1mm} \frac{\partial f}{\partial \phi}g = - \int\limits_0^{2\pi}\mathrm{d}\phi\hspace{1mm} f\frac{\partial g}{\partial \phi},
\end{equation}
with $f$ and $g$ being components of $\hat{\textbf{E}}_n$ and $\hat{\textbf{H}}_n$, respectively, we arrive after some algebra at Eq.~(\ref{Ln}).

\section*{Funding}

We acknowledge support from DFG SPP 1839 and MWK Baden-W\"urttemberg.



\end{document}